# X-ray Variability and Period Determinations in the Eclipsing Polar DP Leo

Craig R. Robinson and France A. Córdova[1]

Department of Astronomy and Astrophysics,
The Pennsylvania State University, 525 Davey Laboratory, University Park, PA 16802
E-mail: robinson@astro.psu.edu, f_cordova@admingw.hq.nasa.gov

## ABSTRACT

We present an analysis of our ROSAT PSPC observations on the eclipsing magnetic cataclysmic variable DP Leo. The soft X-ray spectrum is modeled by a blackbody of kT = $24.8^{+2.6}_{-8.1}$ eV. Severe limits are placed upon the flux from any hard bremsstrahlung component. A strong soft X-ray excess, with respect to hard X-ray emission, is found. The soft X-ray blackbody luminosity is larger than both the cyclotron and bremsstrahlung luminosities.

An upper limit of 500 pc is obtained for the system's distance based upon the X-ray absorption ($N_H$ < 5 x $10^{19}$ cm$^{-2}$) and an estimate of $260^{+150}_{-100}$ pc is determined from a published measurement of the secondary's flux. For the derived blackbody fit, the bolometric luminosity is found to be $L_{bb,bol}$ = $1.4^{+7.1}_{-0.3}$ x $10^{31}$ (d/260pc)$^2$ erg s$^{-1}$.

Absorption by the accretion stream produces an intensity dip prior to each eclipse. Extreme variability in the shape of the light curve from eclipse to eclipse demonstrates that changes in the rate of accretion onto the white dwarf, the sizes of accretion filaments, or variations in the location or amount of absorbing matter in the system occur on time scales shorter than the orbital period (89.8 min). No evidence exists for accretion onto the stronger (59 MG) magnetic pole in the ROSAT data.

A new ephemeris is presented for the eclipse of the white dwarf emission region by the secondary star and another is produced for the orbital conjunction of the two components. The rotation of the white dwarf is shown to be faster than the orbital period by (5.3 ± 1.1) x $10^{-3}$ sec. The origin of the asynchronous rotation may be activity cycle induced orbital period variations or oscillations of the white dwarf's main pole about an equilibrium position. The accretion stream is modeled assuming that disruption of the stream along magnetic field lines occurs close to the white dwarf. The ROSAT intensity dip is explained using this model and the results are shown to be consistent with contemporaneous optical observations. [Note: An earlier paper version of this preprint contained an incorrect value for the cyclotron luminosity.]

*Subject headings:* binaries: eclipsing — stars: individual(DP Leo) — stars: magnetic fields — stars: novae,cataclysmic variables — X-rays: stars

---

[1] Present address: NASA Headquarters, 300 E Street SW, Washington, DC 20546

## 1. Introduction

The eclipsing cataclysmic binary DP Leo belongs to the class of magnetic systems known as either polars, due to their strong optical polarization, or AM Her stars, after a prominent member of the group. These cataclysmic variables (CVs) consist of a synchronously rotating white dwarf primary, containing magnetic fields with strengths of tens of MG, accreting material from a low mass secondary. The magnetic field of the primary is sufficiently strong to disrupt the formation of an accretion disk and instead the accretion flow is directed toward one or more of the white dwarf magnetic poles. However, unlike all other known polars, the accretion stream in DP Leo has been observed to impact the white dwarf up to $\sim 20°$ in longitude behind, in rotation, the line connecting the centers of the two stellar components.

The DP Leo system, originally named E1114+182 after its serendipitous discovery with the EINSTEIN satellite (Biermann et al. 1985), was found during an observation of the nearby NGC3607 group of galaxies. It was the first discovered eclipsing polar and still exists as the shortest period eclipsing polar known (P = 89.8 min). The relative faintness of the system in visual light (V$\gtrsim$18) and lack of hard X-ray emission has hampered investigation. The comprehensive paper by Biermann et al. (1985) established DP Leo as a soft X-ray source with modulations in the X-ray light curve produced by projection effects, as the emission regions rotate with the white dwarf surface, and occultations, when the regions rotate over the white dwarf limb. However, it was not possible to constrain the temperature of the X-ray emitting region from the EINSTEIN data.

The soft X-ray emission in polars is thought to arise through the reprocessing of hard X-ray bremsstrahlung and optical/IR cyclotron emission by the photosphere of the white dwarf (Lamb & Masters 1979), or through the thermalization of large filaments, or "blobs", in the accretion stream which penetrate through the shock region, above the white dwarf's surface, and into its photosphere (Frank, King & Lasota 1988). X-ray observations of DP Leo were additionally obtained with the EXOSAT satellite in 1984 by Schaaf et al. (1987) who reported a lack of detection in both the argon chambers of the Medium Energy (1 - 20 keV) experiment and with the aluminum/parylene filter on the low energy Channel Multiplier Array. The Schaaf et al. research note does provide a noisy light curve for DP Leo using the 3000Å("thin") lexan filter, which has a greater efficiency at very soft energies ($\sim 0.1$ keV) for soft blackbody emission than the Al/P filter. This light curve exhibits an eclipse consistent with the Biermann et al. ephemeris and provides some evidence for a dip prior to eclipse and two pole X-ray emission.

A detailed analysis of the eclipse in red and blue broadband photometry from observations in 1985 was performed by Schmidt (1988). Schmidt concluded from his observations that the cyclotron emission region likely originates in a long accretion arc stretching around $10^9$ cm, up to 60° in longitude, on the surface of the white dwarf. The modeling of cyclotron humps in phase resolved spectra from 1988 to 1990 by Cropper & Wickramasinghe (1993, hereafter CW93) has shown two cyclotron emitting regions to exist at that time. The main emitting magnetic pole was modeled by CW93 using a field of 30.5 MG. This pole was located between 30° and 40° below the rotational equator and found to trail slightly the line of centers between the two stars. The second emission region has a magnetic field strength of 59 MG, nearly twice that of the main pole, and is positioned near the upper rotational pole but nearly opposite to the secondary star (see Figure 1). The differing magnetic field strengths of the two poles and their angular separation of significantly less than 180° implies a magnetic field configuration more complex than a centered dipole.

Recently, a period of 8 months produced a great increase in the number of observations of the DP Leo system. Data were obtained using the Hubble Space Telescope (HST) in 1991 October by Stockman et al. (1994), ground-based observations between 1991 October and 1992 April by Stockman et al. (1994) and Bailey et al. (1993), and ROSAT Position Sensitive Proportional Counter (PSPC) observations from 1992 May 30 to June 1 reported in this paper.

On 1991 October 31, time resolved ultraviolet spectroscopy was performed using the Faint Object Spectrograph on HST by Stockman et al. (1994). Two UV emitting components were observed and identified as the photosphere of the white dwarf (T $\sim$ 16,000K) and a region near the main magnetic pole emitting like a blackbody of around 50,000K. These observations were supplemented by optical spectroscopy 9 days before and spectropolarimetry approximately one month after the HST observations. These optical data confirm that the system was in a "low state", believed to be produced by a reduction in the mass accretion rate in such systems, during this time period. The size of the UV blackbody emission region is similar in size to the cyclotron emission region deduced by CW93 and much smaller than the accretion arcs derived by Schmidt(1988). Further observations by Stockman et al. were obtained using optical spectroscopy on 1992 April 8 showing that the system had again returned to its high accretion state.

Photometric observations of DP Leo by Bailey et al. (1993, hereafter BWF93) on 1992 March 5 showed that the system had returned to its high state by this time as well. They suggest that the longitude of the accretion stream impact spot in DP Leo has changed from $-14° \pm 4°$ (1984) to $+4° \pm 3°$ (1992), where a positive accretion spot longitude implies that the spot leads the secondary component. Furthermore, BWF93 have determined physical parameters by modeling their light curves and show the accretion region to be approximately 10° below the white dwarf equator.

In this paper, we analyze ROSAT PSPC observations of DP Leo and the relationship of the soft X-ray emission to emission at other frequencies. The lack of hard X-ray photons in this system required the use of the ROSAT satellite to obtain X-ray data. Our results provide evidence for a change in the accretion stream geometry since the earlier X-ray observations with EINSTEIN. We develop a model consistent with both our X-ray data and contemporaneous optical data for an accretion stream which is diverted by the white dwarf's magnetic field close to the white dwarf surface.

Our observed soft X-ray flux is too large to have as its origin the reprocessing of bremsstrahlung emission. Furthermore, the large variations observed suggest that reprocessed cyclotron is also not a likely origin for the soft X-rays. Instead, models assuming that accretion occurs in clumps appear to be more consistent with our results.

Eclipse, orbital and rotational periods are determined for DP Leo. Deviations of the white dwarf rotational period from the orbital period may be explained by cyclical variations of the orbital period or oscillations of the main magnetic pole on the white dwarf about its equilibrium position. A preliminary analysis of the work presented here, and on another cataclysmic variable, TT Ari, may be found in Robinson & Córdova (1994a; hereafter Paper I). An additional report on DP Leo was presented in Robinson & Córdova (1994b).

## 2. Observations

ROSAT PSPC observations were obtained of DP Leo during four intervals between 1992 May 30 and June 1 totaling 8910 seconds. The observation intervals covered three eclipses of the X-ray emitting region by the secondary. The orbital period of DP Leo ($\sim$ 89.8 min) is only slightly shorter than the orbital period of the ROSAT spacecraft (96 min) allowing the observation of two consecutive eclipses. Another eclipse was observed one day earlier. Photons detected within the acceptable time windows were extracted from a circular region centered on DP Leo with a radius of 3.5 arcminutes. Five nearby regions, each with the same radius as the source region and apparently free of sources, were used to determine the background level. The background subtracted source count rate was $0.27 \pm 0.01$ cts s$^{-1}$ over the entire observation and $0.34 \pm 0.01$ cts s$^{-1}$ within 0.27 phase of the center of eclipse (the "bright phase").

## 3. Spectral Analysis

### 3.1. Detector Response

The spectral analysis was performed using the "DRM 36" response matrix which is the appropriate choice for ROSAT AO2 observations (Briel et al. 1993). The background subtracted source spectrum was binned from 256 energy channels into the Standard Analysis Software System (SASS) 34 energy channels. The highest 2 of the 34 energy channels were excluded from further analysis due to uncertainties in the detector response. The lowest 4 energy channels are known to be problematic due to the incorrect position determination of some low energy X-ray photons producing "electronic ghost images" (Nousek & Lesser 1993). The effect of the soft energy halo created by these ghost images is greatest in the first two energy channels and was expected to be minimized in our analysis by extracting the source using a large radius. However, a drift in the energy to PI (pulse invariant) channel transformation, calibrated at a single energy of 1.49 keV but found to vary non-linearly with energy, may have caused a deficit of photons in the lowest two energy channels (Briel et al. 1993). This effect is seen in the ROSAT spectrum of DP Leo (Figure 2) where the steep initial increase in the spectrum, visible in the first 3 channels, is larger than is plausible for the energy resolution of the PSPC detector at these low energies. Therefore, even though the DP Leo spectrum is quite soft with no significant detection of source photons in energy channels 12 or above (E > 0.47 keV), we determined it necessary to exclude the first 2 energy channels (E < 0.11 keV) from spectral fits. This limitation does not severely restrict the ability of the ROSAT observations to constrain column densities or blackbody spectral fits.

### 3.2. Soft X-ray Spectrum

The soft X-ray component in the spectra of polars may arise from large accretion column filaments which permeate the shock region above the white dwarf and reach the photosphere where they cool by emitting thermalized soft X-rays (Frank et al. 1988). An alternative suggestion is that reprocessing in the white dwarf photosphere of hard X-ray bremsstrahlung emission, produced in the post-shock region, and cyclotron radiation, generated by free electrons flowing along the magnetic field lines, produces the observed soft X-rays. The DP Leo soft spectrum was fit using single component models with absorption. Acceptable fits were obtained both with a power law and with a blackbody component. However, the physical models for the origin of the soft X-rays predict blackbody-like emission to occur. For the power law model, the best fit was found using an unabsorbed power law with spectral index, $\alpha$, of 4.0. The resulting $\chi^2$ statistic for this model was 23.5 (28 degrees of freedom). The best fit blackbody model was produced with kT=$24.8^{+1.9}_{-5.4}$ eV and $\chi^2$ = 25.3 (28 d.o.f.), where again the source was unabsorbed (see Figure 2). The calculation of 90% confidence intervals for the blackbody model limits the absorption column density to $N_H$ < 5 x $10^{19}$ cm$^{-2}$ (see Figure 3).

EINSTEIN observatory observations of DP Leo from 1979 (see Biermann et al. 1985) were extracted and a spectral fit performed to the data. The spectral models were not well constrained due to the low sensitivity of EINSTEIN at soft energies. However, the data were consistent with the ROSAT blackbody spectral fit and a harder component was not found. EXOSAT observations of DP Leo were obtained in 1984 with a firm detection in the low energy (LE) detector using the thin lexan filter but no source was detected using the aluminum/parylene filter or with the medium energy (ME) instrument (Schaaf et al. 1987), indicating a low hard X-ray flux.

### 3.3. Harder Component

The number of photons above 0.47 keV possibly corresponding to a harder component is difficult to determine due to the presence of several hard X-ray sources surrounding DP Leo. These emission regions include the diffuse hot gas component around the early-type galaxy NGC3607 and the presence of several point sources. In particular, our spatial analysis of the observed hard

X-ray photons suggests that the greatest contribution to the background, within 3.5 arcminutes of the center of the distribution of photons from DP Leo, are 2 weak sources each centered around 2 arcminutes away from DP Leo. Therefore, we adapted our process of source and background computation to the specific problems of the additional hard sources and the behavior of hard X-ray photons in the PSPC detector.

Source photons between 0.47 and 2.28 keV were extracted from a region 1.36 arcminutes in size surrounding the center position of DP Leo. The expected encircled fraction of hard X-ray photons from the source falling within this region ranges from 90% at 1.7 keV to 97% at 0.5 keV (Hasinger et al. 1992) for the PSPC detector. The contribution from the two weak nearby sources to this extracted region is expected to be less than 1 photon. The background region selected for comparison was an annulus surrounding DP Leo with inner radius 3.0 arcminutes and outer radius of 3.75 arcminutes. The region within 45° of increasing declination was discarded from the background calculation due to the presence of nearby sources. The chosen background region samples nearly equally regions closer and further from the cluster gas yielding a reasonable estimate for the contamination in the source region. A total of 25 photons were detected in the source region and $7.8 \pm 2.8$ photons in the background region when scaled to the source region area. The 25 detected source photons were scaled to 26.7 photons after correction for the number of photons expected to have fallen outside the extracted region of 1.36 arcminutes.

Upper and lower limits to the source intensity contributing the observed counts were calculated following the Bayesian statistical method (Kraft, Burrows & Nousek 1991). The 90% confidence interval upper bound is 28.4 counts and lower bound is 11.3 counts. This results in a maximum count rate of $3.2 \times 10^{-3}$ cts s$^{-1}$ and minimum count rate of $1.3 \times 10^{-3}$ cts s$^{-1}$ attributable to the source for photons between 0.47 and 2.28 keV. Hard X-ray components are clearly observed in other polars with similar magnetic field strengths, implying perhaps that the rate of the shock-producing accretion per unit area is lower in DP Leo than in these other systems.

## 4. Distance

The only distance estimate to DP Leo was a lower limit of 380 pc by Biermann et al. (1985). They determined this lower limit using their mass estimate for the secondary of $0.098 M_\odot$ and their non-detection of the secondary to an R magnitude of 22. The compilation of low mass star absolute magnitudes in Young & Schneider (1981, hereafter YS81) was used by Biermann et al. to estimate the absolute R magnitude for the secondary. This resulted in their lower limit estimate of the distance. A measurement of the R magnitude of the secondary and its mass was recently performed by BWF93 from a fit to their photometric light curve. They obtained a solution with a mass of 0.106 $M_\odot$ for the secondary and an R magnitude of 21.8. In Paper I, we re-evaluated the distance to the DP Leo system using the same method as Biermann et al. to determine the distance to DP Leo based upon the tables of YS81. However, more recent work on the physical parameters of low mass stars (see Bessell & Stringfellow (1993) and Burrows & Liebert (1993) for reviews) has shown that the (V-R) colors in YS81 are too red for a given spectral class.

An empirical M-L relationship for low mass stars was derived by Henry (1991). For a $0.106 M_\odot$ star, the absolute K magnitude, $M_K$, was found to be $9.25 \pm 0.40$ which corresponds to a spectral type of M6V in the system of Bessell (1991) for old disk M dwarfs. This also agrees with the results of YS81 for the value of $M_K$ and the spectral classification of M6V, for a star of this mass. Yet YS81 give a (V-R) color index of $+2.54$ and $M_V = 16.08$, implying $M_R = 13.54$, while Bessell (1991) determined a value for (V-R) of $+1.91$ and $M_V = 16.63$, yielding $M_R = 14.72$. Using the newer estimate by Bessell, we estimate the absolute R magnitude for the secondary in the DP Leo system to be $14.7 \pm 1.0$ mag which results in a distance to the system of $260^{+150}_{-100}$ pc. This is lower than our estimate in Paper I.

An independent upper limit to the distance is obtained from the X-ray spectral fit. The DP Leo system is located in a region of unusually low equivalent hydrogen column density. The distance could be as far as approximately 500 pc, based upon the observations of this region by Frisch & York (1983), while still exhibiting a column density $< 5 \times 10^{19}$ cm$^{-2}$ (the upper limit in our 90% confidence contour).

## 5. Luminosity

Polars often exhibit a strong excess of soft X-rays relative to the flux of a harder bremsstrahlung continuum. The radiative models of soft X-ray emission in polars suggest that the observed soft X-rays are reprocessed bremsstrahlung and cyclotron emission by the photosphere of the white dwarf. Therefore, depending upon the albedo, the sum of the bremsstrahlung and cyclotron luminosities, $L_{br} + L_{cyc}$, should be roughly equal to the blackbody luminosity, $L_{bb}$. However, in many polars, it has been found that $L_{bb} \gg (L_{br} + L_{cyc})$ producing the "soft X-ray excess problem". Also, in many polars, bremsstrahlung cooling dominates over cyclotron cooling and thus the contribution of $L_{cyc}$ is often neglected leaving only the blackbody and bremsstrahlung luminosities to be compared.

The observed soft X-ray flux during the DP Leo bright phase was measured to be $F_{bb}(0.1\text{-}0.5 \text{ keV}) = 1.3 \times 10^{-12}$ erg cm$^{-2}$ s$^{-1}$. It was necessary to deconvolve the detector response and the observed counts per channel using an assumed spectral shape, in this case our blackbody fit, to obtain this result. The behavior of the ROSAT light curve, especially the portion after eclipse, appears consistent with an origin from dense blobs or filaments which produce protruding splashes that radiate isotropically into half space (Hameury & King 1988, Beuermann & Schwope 1989). Therefore, here we assume that $L_{bb} = 2\pi d^2 F_{bb}$ which yields $L_{bb} = 5.4 \times 10^{30}$ (d/260pc)$^2$ erg s$^{-1}$, for photons between 0.1 and 0.5 keV. The bolometric blackbody luminosity results in a value of $L_{bb,bol} = 1.4^{+7.1}_{-0.3} \times 10^{31}$ (d/260pc)$^2$ erg s$^{-1}$ where the range in luminosity is determined by the 90% confidence contour of the spectral fit.

An estimate of the flux from a harder bremsstrahlung component is determined by fitting a 30

keV bremsstrahlung component to the few harder counts observed. This fit to the observed data yields a flux at the Earth of $F_{br,bol} = 2 \times 10^{-13}$ erg cm$^{-2}$ s$^{-1}$ and a bolometric bremsstrahlung luminosity of $L_{br,bol} = 2\pi d^2 F_{br,bol} = 8^{+30}_{-8} \times 10^{29}$ (d/260pc)$^2$ erg s$^{-1}$. The flux resulting from the best fit spectral model falls within the count rate limits, attributable to the source, obtained using the Bayesian statistical approach (§3.3).

The cyclotron luminosity from DP Leo was estimated by Ramsay et al. (1994) using the photometry of BWF. Their resulting value of $L_{cyc} = 1.1 \times 10^{30}$ (d/260pc)$^2$ erg s$^{-1}$ is an order of magnitude lower than that of the soft X-ray blackbody, but similar to the estimate of the bremsstrahlung emission.

The area of the X-ray blackbody emission may be deduced from the blackbody spectral fit and was found to be $A_{X,bb} = 4 \times 10^{13}$ (d/260pc)$^2$ cm$^2$. Based upon the white dwarf size derived by BWF93, the fraction of the white dwarf surface emitting these X-rays is found to be $f_{X,bb} = 5 \times 10^{-6}$ (d/260pc)$^2$. The cyclotron emission region (BWF93) and UV emission spot (Stockman et al. 1994) correspond to emission from a larger region: $f_{cyc} \simeq f_{UV,bb} \simeq 6 \times 10^{-3}$. Therefore, for reasonable estimates of the distance, we find that $f_{X,bb} \ll f_{UV,bb} \simeq f_{cyc}$.

During the low state in 1991 October, Stockman et al. (1994) measured the best fit blackbody luminosity extrapolated over EUV wavelengths to be $L_{EUV,bb} = 1.2 \times 10^{31}$ (d/260pc)$^2$ erg s$^{-1}$. At the time of those observations, the R magnitude at maximum light, $R_{max}$, was around 18.2 as compared to $R_{max} \sim 16.9$ during the high state of 1992 April (Stockman et al. 1994). Therefore, in the low state, the flux in R dropped by a factor of around 3.3. Assuming a constant value for the electron temperature and optical depth parameter, the cyclotron luminosity during the low state is estimated to be $L_{cyc,low} \simeq 3.5 \times 10^{29}$ $(d/260pc)^2$ erg s$^{-1}$. Thus it is found that $L_{EUV,bb} \gg L_{cyc,low}$. The areas and luminosities of the cyclotron and UV blackbody emission regions are similar, but the much larger UV blackbody emission implies that its origin is not from reprocessed cyclotron emission.

The observed luminosity in the high state is obtained as the sum of the individually observed components: $L_{obs,tot} = L_{bb,bol} + L_{br,bol} + L_{cyc} = 2 \times 10^{31}$ (d/260pc)$^2$ erg s$^{-1}$. Using $M_{wd} = 0.71$ M$_\odot$ and $R_{wd} = 0.018a = 7.7 \times 10^8$ cm, as derived by BWF93, together with our distance estimate of 260pc, the resulting mass accretion rate is found to be $\dot{M} = 2 \times 10^{14}$ g s$^{-1}$. Autocorrelations of the after-eclipse data for each orbit show characteristic time scales on the order of $\sim 10$ sec which suggests that the time scale for the tip to end impact of an accretion filament is of order 10 seconds. The mass accretion rate and the approximate upper limit to the total blob accretion time implies that individual filament masses are, on average, $M_{blob} \lesssim 2 \times 10^{15}$ g with linear dimension $R_{fil} \lesssim 6 \times 10^9$ cm $\simeq 8$ $R_{wd}$.

## 6. Light Curve Analysis

### 6.1. Individual ROSAT Orbits

The X-ray light curve from the PSPC, derived using all 34 energy channels, is presented in the top plot of Figure 4 for the entire data set and the lower plots for each of the individual eclipses. The data are binned into 500 equally sized bins of length 10.78 seconds each with the phase calculated based upon the linear ephemeris determined in this paper (§6.2.1). Two dips are visible in the light curves with the first dip centered near phase 0.947. The second dip, at 1.00 phase, is suspected of being the previously observed eclipse of the X-ray emitting region on the white dwarf by the secondary companion. No evidence exists for accretion onto the second pole, as was suggested by EXOSAT observations from 1984 (Schaaf et al. 1987), cyclotron humps in phased resolved spectra from 1988-90 (CW93), and our analysis of EINSTEIN observations from 1979 (§6.4). The lack of X-ray emission prior to phase $0.73 \pm 0.01$ results from the occultation of the X-ray emitting region during the period when it is on the "back side" of the white dwarf.

Hardness ratios are shown on the right side of Figure 4 for each of the ROSAT light curves. The ratio was defined as PI channels 18 through 46 (0.18 - 0.47 keV) divided by PI channels 8 through 17 (0.08 - 0.18 keV) in the 256 channel system. This definition produces a hardness ratio near unity for the entire observing set. For the individual orbits, hardness ratios were determined for the period before and after the center of eclipse. Hardness ratios for the combined data set were calculated over intervals of 0.1 in phase.

The ROSAT light curves show great variability within each orbit and variations in the shape of the light curves from orbit to orbit. However, a power spectrum derived from the X-ray arrival times showed no evidence of strong periodicities or quasi-periodic oscillations on time scales shorter than the orbital period. The first ROSAT observation interval that included coverage of an eclipse, the Eclipse 1 data set in Figure 4, showed little emission prior to the first intensity dip. A spike between the position of the intensity dip and the eclipse was evident and strong variable X-ray emission was observed after the eclipse. The next observation, the Eclipse 17 data set (16 DP Leo orbits later), showed stronger emission before the intensity dip than after the eclipse, and again a spike between dip and eclipse. The final observation, the Eclipse 18 data set (only one DP Leo orbit later), showed emission before the intensity dip similar to the Eclipse 17 set, and flaring behavior after eclipse similar to the Eclipse 1 observations. Data were lost for a brief period of time around eclipse egress in the Eclipse 18 data set.

The hardness ratios determined from the combined data set exhibit a marginally higher value at the start and end of the bright phase. ROSAT observations of another eclipsing polar, UZ For, by Ramsay et al. (1993), suggested that the X-ray emission in the rapid rise to the bright phase was softer than the emission throughout the majority of the bright phase. This was interpreted as evidence for a structured emission region with lower temperatures in the outer regions of the emission spot being the first to come into view. This behavior is not evident in our DP Leo observations.

The count rate prior to the eclipse in the Eclipse 1 data set is far less than the much stronger emission in the Eclipse 17 set. The hardness ratio of these data stayed roughly the same or increased slightly for Eclipse 17 observations as compared to Eclipse 1. A small increase in the absorption column density to $N_H \sim 2 \times 10^{20}$ cm$^{-2}$, caused perhaps by material around the accretion stream or shock region, could account for the lower emission in the Eclipse 1 set and still fall within the tolerances of the hardness ratios. However, inferences based upon hardness ratio variations are limited due to the paucity of observed photons.

### 6.2. New Eclipse and Orbital Ephemerides

#### 6.2.1. Linear Ephemeris

In Figure 5, the combined ROSAT light curve is shown together with the contemporaneous broadband R light curve of BWF93 and the X-ray data from EINSTEIN. The phase, according to prior ephemerides, and duration of the optical eclipse coincide with the second ROSAT dip confirming its origin as the eclipse by the secondary (see Figures 5 and 6). The duration of the ROSAT X-ray eclipse is 3.6±0.3 minutes. Times of eclipse centers were calculated for the two observed eclipses with complete coverage and found to occur at HJD2448773.21442±0.00012 and HJD2448774.21235±0.00012. A revised eclipse ephemeris is then obtained using the above timings together with those listed by Biermann et al. (1985), Schaaf et al. (1987), Schmidt (1988), Bailey (1993), and Stockman et al. (1994) with each timing weighted according to its associated error:

$$\text{Eclipse Center} = \text{HJD}2444214.552929(43) + 0.0623628440(7)E.$$

The resulting uncertainties are given in parentheses for the last significant digits. Figure 7a plots the eclipse timings used in the determination of the linear ephemeris against the deviations from this ephemeris.

#### 6.2.2. Emission Region Longitude Variations

Figure 7b shows the variation in longitude of the emission spot position on the white dwarf according to results from Biermann et al. (1985), BWF93, and Stockman et al. (1994) together with determinations from the EXOSAT and ROSAT observations. EXOSAT observed the start and end of the bright phase and therefore the midpoint of the two yields an estimate of the emission spot longitude. For the EXOSAT data from 1984 June, this resulted in a longitude estimate of $-12° \pm 7°$ and for the data from 1984 December, a longitude estimate of $-11° \pm 7°$.

The DP Leo system has an inclination of around 80° and a cyclotron spot co-latitude in 1992 March of 100° (BWF93). The start of the X-ray bright phase at $\phi = 0.73 \pm 0.01$ is consistent with the red data of BWF93 which started its bright phase at 0.73 and ended at phase 0.25. The end of the bright phase was not covered by our ROSAT observations. The X-ray ingress and egress from eclipse are rapid and only a large upper limit of 0.004 phase ($\sim$ 22 sec) can be placed upon the duration. Since the eclipse of the white dwarf photosphere takes around 51 seconds, this does not yield a stringent limit for the size of any accretion region. If the latitude and size of the X-ray emission region is approximately the same as that of the cyclotron region, then a similar estimate for the accretion spot longitude is obtained for a region at longitude $+4° \pm 6°$. The error is estimated from the uncertainty in the start of the X-ray bright phase and from expected ranges for the size and location of the X-ray emission region.

A linear fit to the longitude variations in Figure 7b is plotted as a dashed line and implies a progression of $2°.05 \pm 0°.42$ per year for the longitude of the spot. Variations occur in the time of eclipse due to the movement of the emission region relative to the position of the secondary. The accretion spot longitude does not appear to be a function of accretion rate since the low state observations by Stockman et al. do not deviate from the linear fit. The changing longitudes are likely produced by variations in the position of the accreting magnetic pole on the surface of the white dwarf relative to the position of the secondary. If the accreting magnetic pole is coupled to the white dwarf surface, then the white dwarf rotation period is slightly shorter, by $(5.3 \pm 1.1) \times 10^{-3}$ sec, than the orbital period. This is a deviation of less than one part in $10^6$:

$$\frac{P_{orb} - P_{rot}}{P_{orb}} = (9.7 \pm 0.2) \times 10^{-7}.$$

The changing longitude of the emission region on the surface of the white dwarf also implies that the observed eclipse period is slightly longer than the actual orbital period of the system. When the emission region moves in the positive longitude direction (i.e., in the direction of rotation) at a given latitude, the secondary must move further in its orbit to eclipse the emission region. The amount of additional time required depends upon the rate of the longitudinal shift and position of the spot at the time. We compute the additional time required to reach eclipse center for the 1979 to 1992 time period using our fit to the emission region variations in longitude. For a region near the rotational equator, a relationship between the observed eclipse period, $P_{ecl}$, and the true orbital period, $P_{orb}$, is obtained as:

$$\frac{P_{ecl} - P_{orb}}{P_{orb}} = (1.7 \pm 0.4) \times 10^{-8}.$$

Therefore, we find that $P_{ecl} > P_{orb} > P_{rot}$.

### 6.2.3. Eclipse, Orbital and Rotational Periods

The eclipse period and observed shifts in longitude yield the information necessary to calculate the orbital period of the system and rotation period of the white dwarf. Using the linear ephemeris previously obtained, the eclipse, orbital and rotational periods are

$$P_{ecl} = 0^d.0623628440(7),$$

$$P_{orb} = 0\overset{d}{.}0623628429(8), \text{and}$$

$$P_{rot} = 0\overset{d}{.}0623627824(14).$$

The fit to the variations in longitude predicts the crossing of 0° longitude by the emission region to have occurred in the year 1990.4±2.9. This information allows an orbital ephemeris of the system to be derived:

$$\text{Orbital Conjunction} = \text{HJD}2448038.455435(88) + 0.0623628429(8)\text{E}$$

where "Orbital Conjunction" is the time when the secondary is at inferior conjunction and the error in the time of conjunction includes the uncertainty in the meridian crossing.

However, there is no reason to believe that the variation in emission longitude position is strictly a linear function with time. It has been suggested by both Wickramasinghe & Wu(1991) and King & Whitehurst (1991) that oscillations with periods of around 30 to 50 years are possible in the orientations of the magnetic fields of the white dwarf with respect to the position of the secondary. These mechanism could be producing the observed shifts in accretion longitude, but another possibility is that oscillations in the orbital period result in accretion longitude variations.

### 6.2.4. Variations in Orbital Period

As recently discussed in Stockman et al. (1994), a negative time derivative to the period will better fit the observed eclipse timings. A second order polynomial fit to the eclipse timings is shown as a dashed line in Figure 7a and results in the following quadratic ephemeris:

$$\text{Eclipse Center} = \text{HJD}2444214.552726(94) + 0.062362855(49)\text{E} - 1.27(52)\text{x}10^{-13}\text{E}^2.$$

The necessity of the second order term was determined using the $F_\chi$ statistic (Bevington & Robinson 1992). There exists a 99.7% probability that the second order term is required in this model.

The time derivative of the period is found to be $\dot{P}_{orb}$ = (-4.1 ± 1.7) x $10^{-12}$ s s$^{-1}$ which gives a characteristic time of $P_{orb}/\dot{P}_{orb}$ = (-4.2 ± 1.7) x $10^7$ yr, much shorter than the time expected due to gravitational radiation alone ($\sim$ 3 x $10^9$ yr, see Wickramasinghe & Wu 1994). Stockman et al. suggest that the sign of the time derivative implies that the secondary, which is losing mass to the white dwarf, is still in thermal equilibrium. However, the orbital period, which is decreasing, is still larger than the rotational period. Instead, the period change may originate from magnetically induced variations in the rotational oblateness of the secondary star. These variations alter the secondary's gravitational quadrupole moment and produce changes in the orbital period of the system (Applegate 1992).

Quasi-periodic orbital variations in close binary systems containing low mass main sequence stars have been found for various types of binaries including Algols (Bolton 1989), RS CVn

– 13 –

systems (Hall 1989), pulsar-low mass star binaries (Arzoumanian, Fruchter & Taylor 1994) and cataclysmic variables (Warner 1988). The cycle lengths for the period variations in each of these systems ranges from a few years to several decades. The monitoring of chromospheric activity in 111 lower main sequence stars by Baliunas et al. (1994) has produced evidence for cyclic variations analogous to the 11 yr solar cycle in many stars in their sample. These cycle periods were determined to extend over a similar range to the orbital period variations in the binary systems (from 2.5 yr to beyond 25 yr).

### 6.3. Application of Rotational Angular Momentum Changes to DP Leo

The secondary star in DP Leo likely rotates synchronously, or nearly synchronously, with respect to the orbit implying an equatorial rotational velocity of around 110 km s$^{-1}$. This is much faster than the average rotational velocity of less than 2 km s$^{-1}$ for M dwarfs in the solar neighborhood (Marcy & Chen 1992). Magnetic activity is correlated with rotation rate in single lower main sequence stars implying that the secondary in DP Leo may exhibit a high level of magnetic activity. However, the secondary also has a low mass ($M_{sec} = 0.106 M_\odot$) which suggests that the star is completely convective and argues against the conventional model of a magnetic dynamo existing in this star. Nevertheless, Fleming et al. (1993) have observed X-rays in stars of lower mass than this, down to spectral type M8, suggesting that a dynamo process may still be active in stars at very low masses. The relatively large distance to the DP Leo system and emission from the region around the primary star has hindered direct observational evidence of this magnetic activity from the secondary.

Applegate (1992) proposed a mechanism to produce the observed changes of orbital period in close binaries through variations in the distribution of rotational angular momentum in the magnetically active star. During the portion of high magnetic activity in a star's magnetic cycle, angular momentum is transferred outward from the center of the star producing a star more closely in solid body rotation. The outer layers of the star will spin up, producing a more oblate outer region, while the interior spins down and becomes less oblate. A larger gravitational quadrupole moment then exists since the outer oblate region, farther from the spin axis than the spun down interior region, dominates the change in the quadrupole. The result is a decreasing orbital period as observed in DP Leo. During times of low magnetic activity, the star is driven away from solid body rotation, and the rotational kinetic energy is higher than it was in the high activity state. In this case, an increasing orbital period results.

The angular momentum transfer required to yield the observed change in period in DP Leo is given by Applegate (1992) as

$$\Delta J = -\frac{GM_{sec}^2}{R_{sec}} \left(\frac{a}{R_{sec}}\right)^2 \frac{\Delta P_{orb}}{6\pi},$$

where $\Delta J$ is the angular momentum transferred, $M_{sec}$ is the mass of the secondary star, a is the semi-major axis of the binary system, $R_{sec}$ is the radius of the secondary, and $\Delta P_{orb}$ is the amplitude of the orbital period change. Using the physical parameters from BWF93 and $\Delta P_{orb}$ = -1.6 x $10^{-3}$ sec, determined from $\dot{P}_{orb}$ over the time period between the EINSTEIN and ROSAT observations, yields $\Delta J$ = 5 x $10^{44}$ g cm$^2$ s$^{-1}$. A relationship between the subsurface magnetic field of the magnetically active star, the physical parameters of the binary star system, the amplitude of the period variations, and the magnetic cycle period is given by Applegate (1992) and modified according to Applegate (1994) to be

$$B_{sub,sec}^2 \sim \frac{GM_{sec}^2}{R_{sec}^4} \left(\frac{a}{R_{sec}}\right)^2 \frac{\Delta P_{orb}}{P_{cyc}},$$

where $B_{sub,sec}$ is the subsurface magnetic field of the secondary star and $P_{cyc}$ is the cycle length for the orbital period variations. From Figure 7a, it is observed that the variations in orbital period, if periodic or quasi-periodic, take place on time scales larger than the baseline of observations (12.5 yr). This is consistent with the time scale of orbital period oscillations determined in several long-observed close binaries (e.g., 32 yr variations in Algol [Söderhjelm 1980]). Since observations only exist over a portion of any longer term cyclic variations, we use $\Delta P_{obs}$, the period change determined above, during the baseline of observations from EINSTEIN to ROSAT, $P_{base}$ = 12.5 yr, to approximate

$$\frac{\Delta P_{orb}}{P_{cyc}} \approx \frac{|\Delta P_{obs}|}{P_{base}}$$

in the previous equation. The subsurface magnetic field of the secondary is then found to have a field strength of $B_{sub,sec} \sim$ 5kG.

Several tests are possible of the application of the model for activity induced period variations to the DP Leo system. For example, the luminosity of the secondary is expected to be higher during the high activity state, by around 14% (Applegate 1994), since the energy previously producing differential rotation is again available to the star. Unfortunately, observing this change is extremely difficult due to the low luminosity of the secondary and the emission from the region around the primary. The interval of high magnetic activity produces an increased oblateness of the secondary and a decreased distance between the stellar components. This should result in a higher mass transfer rate and therefore both the cyclotron and X-ray luminosities should be greater. But occasional rapid drops to low mass transfer states in DP Leo and other polars suggest that shorter time scale mechanisms may also control mass transfer in these systems. This mechanism could be occasional abrupt changes in the secondary, as angular momentum is redistributed, varying the position of the $L_1$ point and therefore the mass transfer rate. Of course, another manifestation of cyclic orbital period variation may be differences between the orbital and rotational periods in the system, as is observed in DP Leo.

If the synchronization time scale, $\tau_{sync}$, for the white dwarf's rotational period to equal the orbital period is large compared to the time scale for orbital period variations, then differences between the white dwarf rotation and the orbital period would be expected. Estimates of the synchronization time scale vary, but estimates range from around $10^4$ to $10^7$ yr (cf., Hameury,

King & Lasota 1989). Our period determinations for DP Leo have found that $P_{orb} > P_{rot}$ with evidence that $\dot{P}_{orb}$ is negative. The orbital period is decreasing but is currently larger than the white dwarf rotation period, a result unexpected if the system is asynchronous due to a constantly decreasing orbital period. At the current rate of orbital period decrease, the two periods will coincide in $\sim 40$ yr.

If, however, the orbital period is oscillating in DP Leo due to a solar-like activity cycle on the secondary star and $P_{cyc} \ll \tau_{sync}$, then the rotation of the white dwarf will stay near the mean of the orbital period while the orbital period fluctuates according to the activity cycle. Just after orbital period maximum and lasting for around one-quarter of the total cycle, the orbital period will be longer than the white dwarf rotational period yet the orbital period will be decreasing as observed in DP Leo. When the orbital period is shorter than the rotational period of the white dwarf, the main accretion region will appear to move in the reverse direction, toward more negative longitudes, as has been observed in the polar WW Hor (BWF93).

### 6.4. Light Curve Modeling

Light curve models for the X-ray emission observed in DP Leo were computed assuming that the emission was produced by an optically thick blackbody modulated by the cosine of the viewing angle. Such a feature could exist from reprocessing or through emission from a dense region of blob accretion. The photometric models assume that the light curve varies due to viewing angle effects and occultation of the emission regions as they rotate out of view. The modeling routines allow the fluxes of two regions as well as their sizes, shapes and positions to be fixed, constrained between limits or allowed to vary freely while the Levenberg-Marquardt method was used to fit the model to the observed X-ray light curves.

The numerous free parameters were limited in the initial modeling of the EINSTEIN data to constrain the X-ray emitting regions to the positions of the cyclotron emission regions found by CW93 (as shown in Figure 1). The relative flux per unit area of the two regions were allowed to vary and the fit to the data is shown in Figure 5. However, numerous other fits to the data are possible depending upon the number of interdependent parameters allowed to vary. Visual inspection of the EINSTEIN data suggested a need for a second X-ray emission region due to the above zero X-ray flux between phases 1.3 and 1.5 (see Figure 5). The requirement of a second emission region as compared to a single arc or spot is quantitatively tested through the use of the $F_\chi$-test. Subtracting the $\chi^2$ values from two and one emission region fits and dividing by the reduced $\chi^2$ value of the two region fit produces an $F_\chi$ statistic large enough to state that the flux from the second region exists at the 90% probability level in this model. The main accretion region dominates the X-ray flux in this solution from phase 0.85 to near phase 1.3 where the secondary emission region begins to be the larger contributor. Optical observations from 1982 to 1989 (Biermann et al. 1985, CW93) show that accretion was occurring at a second pole and support the likelihood of two pole accretion also existing at the time of the EINSTEIN observations.

The X-ray emission in 1992 from ROSAT shows no evidence of X-ray emission at the position previously determined for the second pole according to prior cyclotron and X-ray observations. The cyclotron emission dominating the broadband R photometry from earlier in 1992 was fit by BWF93 with a single circular emission region centered at or near the rotational meridian with a radius of 0.15 times the white dwarf's radius. Our estimate of the X-ray emitting surface area is much smaller than this, but models using one or two regions of emission producing variations in X-rays through projection effects alone cannot produce the observed X-ray variations.

A model consisting of a circular emission region located at the rotational latitude of the cyclotron emission region, $-10°$, determined by BWF93, can reproduce the composite ROSAT light curve prior to the intensity dip. It is, however, a poor representation of the region after eclipse (see Figure 5). It is not possible to constrain the radius of a single soft X-ray emitting spot producing the light curve prior to eclipse other than that it must be less than a radius of several degrees. The lack of X-ray emission outside of the bright phase in the ROSAT data suggests that if more than one emission region were to exist, they would all need to be close together in rotational longitude. The position of different emission regions is additionally constrained by the fact that the drop to background flux in the intensity dip, presumably produced due to occultation by the accretion stream, needs to act upon the entire X-ray emission region at around the same time. A model to explain the observed X-ray light curves using a single emission region that contains a spread in accretion rates is discussed in a later section (§7.1).

The intensity dip lasts for a duration of 3.8±0.3 minutes with a center at 0.947±0.002 phase. The likely origin of the observed dip is obscuration of the X-ray emission region on the surface of the white dwarf by the accretion stream. A limit of $N_H \gtrsim 5 \times 10^{20}$ cm$^{-2}$ is obtained for the column density of the absorbing material based upon the X-ray observations. No similar dip was observed in contemporaneous UV and "white light" observations (Stockman et al. 1994). The "white light" observations of Stockman et al. were centered around 3400Å with a FWHM $\sim$ 1900Å. Free-free absorption would be significant here only for high values of the column density. Therefore, we can place further limits on the column density of $N_H \lesssim 10^{23}$ cm$^{-2}$ which is at the upper end of estimates of accretion column densities in polars. This estimate assumes that the UV emitting region, which is larger than the fraction of the surface emitting soft X-rays, is completely behind the accretion stream. If some sections of the UV emitting region are not behind the accretion stream, the column density of the stream could be larger.

Free-free absorption should have a larger effect upon the R-band light curves than the white light and UV observations. The comparison of the ROSAT and R light curves in Figure 6 shows that the shape of the R-band light curve exhibits little or no absorption by the stream at the phase of the X-ray intensity dip. This places a constrain of $N_H \lesssim 5 \times 10^{22}$ cm$^{-2}$ on the column density of intervening material at the time of the R-band observations, again assuming that a significant fraction of the cyclotron emission region is behind the absorbing material. It should be remembered, however, that variations in the position or density of the accretion stream in DP Leo likely occur on time scales of less than one day, based upon changes observed with ROSAT.

Therefore, simultaneous soft X-ray and infrared observations are necessary to truly constrain densities in the stream for DP Leo.

### 6.5. Origin of Soft X-ray Light Curves

The soft X-ray light curve of the polar AM Her is observed to follow a quasi-sinusoidal shape in its "normal" mode while occasionally taking the form of a square wave in its "anomalous" mode. Hameury & King (1988) showed that the normal behavior is consistent with having numerous accreting blobs producing splashes which shadow each other, while the anomalous mode is consistent with a less dense splash region where individual splashes are seen. In DP Leo, the EINSTEIN data are consistent with normal mode accretion at both poles. However, the ROSAT data exhibit both modes each orbit.

The first part of the ROSAT bright phase exhibits a quasi-sinusoidal shape in both the Eclipse 17 and Eclipse 18 data sets. The lower emission prior to the intensity dip in the Eclipse 1 data may be due to the absorption by a portion of the accretion stream which was in a slightly different position during the first eclipse as compared to a day later. The bright phase data after eclipse, however, appear to show numerous flaring events reminiscent of the square wave-like AM Her anomalous state. Both projection effects and absorption in either the shock region or accretion stream could allow one portion of the soft X-ray emission zone to dominate the light curve at any one time. If the leading edge, the portion ahead in rotation, of the emission region were to have a higher filament accretion rate per unit area as compared to the lagging edge, such a light curve behavior, i.e., quasi-sinusoidal for the bright phase first half and flare domination for the second half, would be possible.

The light curve model of BWF93 produced a result that the R-band spot magnitude at ingress is 20.0 mag while at egress it is 19.7 mag. Such a behavior suggests several possibilities: (1) the "spot" geometry may not be circular, (2) the cyclotron emission region is not uniform, perhaps due to differing accretion rates across the emission region, or (3) absorption is different along the line of sight at ingress and egress. Any of these three possibilities should have an effect upon the soft X-ray emission.

The UV/optical observations of Stockman et al. (1994) suggested the presence of a 5 x $10^4$K hot spot while the observed soft X-ray emission is characterized by a temperature of around 2.9 x $10^5$K (= 24.8 eV). The area of emission for the UV blackbody is also much greater than that of the soft X-ray region. However, the UV/optical observations were obtained with DP Leo in a low state when the accretion rate was likely much lower. Therefore, it is unclear whether a lower temperature, larger emitting area UV blackbody still existed at the time of our observations of a soft X-ray blackbody.

The cyclotron luminosity is insufficient to expect that reprocessing in the white dwarf atmosphere has produced the soft X-ray luminosity observed. In addition, the flaring visible in the soft X-rays is not observed in the cyclotron dominated light curves. In Figure 5, the red (R-band) light curve of DP Leo from BWF93 is shown together with the ROSAT and EINSTEIN light curves. The shape of the red curve varies with phase due to the angle of the cyclotron emission to the line of sight. The cyclotron emission peaks at the two phases when the region is nearly perpendicular to the line of sight. The observed X-ray emission appears to peak at a similar phase to the cyclotron emission after eclipse, but that part of the ROSAT light curve was obtained only during one orbit, Eclipse 18 observations, where flare events appear to dominate. The drastic variations in ROSAT X-ray light curve shape and short time-scale variability observed in DP Leo suggests that the majority of soft X-rays are not from reprocessed cyclotron emission but rather from the accretion of individual filaments.

## 7. Accretion Geometry

### 7.1. Emission Region

The structure of the soft X-ray emission region in DP Leo appears to be quite complex. However, it is possible to make some determinations about the structure of the region. The total area of emission is quite small at any instant and likely originates through blob accretion. The time for individual accreted filaments to thermalize in the white dwarf photosphere is small ($\lesssim 0.3$ sec). A larger cyclotron emission region is known to exist, and a UV emission region of similar size and position to the cyclotron emission has been observed. Soft X-ray emission may occur from distinct points around blob impacts at any particular instant allowing for a relatively small soft X-ray emitting surface fraction relative to the cyclotron and UV emission regions. However, the total region over which blobs impact the white dwarf may be as large as the cyclotron and UV emission regions.

The soft X-ray light curves suggest that individual flaring events dominate after eclipse while smooth variations or low emission dominate before eclipse. To explain the relatively smooth light curve prior to eclipse, numerous accretion blobs are required to be accreting at any instant in time. Splashes from the individual blobs will shadow other blobs averaging out the light curve. Therefore, variations in the projected area of the emission region with respect to the line of sight will dominate the light curve shape producing the observed variations.

The portion of the light curve after eclipse does not appear to follow a sinusoidal shape but rather shows abrupt variations in X-ray flux. This behavior, similar to the anomalous state in AM Her (see Hameury & King 1988), may be produced by a sparse accretion region with a lower number of blobs accreting at any instant per unit area as compared with the emission region dominating prior to eclipse. The individual blob events are observed in such a region producing rapid rise and decay patterns in the light curve.

The projected accretion region surface area varies as the white dwarf rotates. If the total region where soft X-ray emission may occur with time is slightly elongated in longitude, then the leading part of the emission region, in rotation, can have a much larger influence on emission at phases prior to eclipse as compared with phases after eclipse. The lagging part of the emission, in longitude, will have its largest influence after eclipse. Therefore, the leading region may have a denser accretion filament rate per unit area than the lagging region thereby producing the smooth variations prior to eclipse and flare-like emission after eclipse. One would thus expect an increase of flare-like activity with phase during the bright portion of the light curve. This is the behavior seen in the combined ROSAT light curve (see Figure 5). The origin of a spread in blob accretion rate may be an accretion stream impacting the the white dwarf surface obliquely or through the motion of the advancing longitude of the emission zone which slowly forces the stream to alter its position.

## 7.2. Accretion Stream

The shape of the accretion stream must be such that it completely blocks the X-ray emission region surrounding phase 0.947 and allows the region to again be visible just prior to eclipse. Also, the emission region on the white dwarf is required to be close to the line of centers (near $+4°$ longitude).

It was BWF93 who pointed out the drastic variations in accretion spot longitude in DP Leo. The study by BWF93 also concluded that the emission region in the polar WW Hor has moved from leading the secondary to lagging the secondary. If the motion of these emission regions results from steady non-synchronous rotation of the white dwarf primary, then the white dwarf must be spinning slower than the orbital period in WW Hor and faster than the orbit in DP Leo. However, as pointed out by BWF93, Cropper (1988) has found that accretion spot longitudes cluster near to $+20°$ in polars. Such a skewed distribution should not occur if steady non-synchronous rotation is common. Orbital period variations as an origin for the accretion longitude changes in DP Leo has already been discussed. However, other physical mechanisms for accretion spot position variations are possible.

A combined centered dipole and inclined quadrupole field for the white dwarf may allow for such a geometry and explain some of the observed features in DP Leo. The multipole field model of polars by Wu & Wickramasinghe (1993) provides for three magnetic poles on the surface of the white dwarf with two narrow poles lying close to the rotational axis and a band-like pole, of field strength half the axial poles, near the rotational equator. Interaction of this field with the magnetic field of the secondary was shown by Wu & Wickramasinghe to provide a magnetic torque strong enough to counteract the accretion torque. A parallel decentered dipole, however, would not provide a sufficient balance to the accretion torque. In addition, the impulse provided through variations in accretion rate could produce small oscillations of the quadrupole axis with periods of order 50 years (Wu & Wickramasinghe 1991). Such a behavior would be observable as variations in accretion longitude as is observed in DP Leo.

The quadrupole field of the white dwarf, which falls off by the distance to the fourth power, channels the accretion stream onto the equatorial pole only near to the white dwarf (within several white dwarf radii). The accretion stream, outside of the range dominated by the quadrupole field, has a motion dictated by the momentum of the Roche lobe overflowing material and the dipole field of the white dwarf. Close to the white dwarf, the quadrupolar field dominates and distorts the stream to the main accretion region.

The main accretion region was determined to be around 10° below the orbital plane by BWF93. Therefore, we modeled the system assuming that the white dwarf has a quadrupole field inclined by 10° tipped along the direction of the line of centers. The physical parameters used in the model were those determined by the photometric fit of BWF93: i=79°.57, $R_{wd}/a = 0.018$, $M_{wd} = 0.71 M_\odot$ and $M_{sec} = 0.106 M_\odot$.

The system is shown to scale in Figure 8. A stream is assumed to flow through the inner Lagrangian point and follow a ballistic trajectory (Lubow & Shu 1975) until it is disrupted by the quadrupolar field where it follows along magnetic field lines. The width of the stream at the disruption point is modeled according to the prescription of Wu & Wickramasinghe (1991). Simple pressure balancing arguments and the work of Wu & Wickramasinghe (1991, 1993) suggest that the stream should be disrupted around 5-10 $R_{wd}$ from the surface of the white dwarf for a quadrupole of field strength around 60 MG, the approximate strength in a dipole/quadrupole field model required to explain the strengths of the observed magnetic poles. Figure 8 is shown assuming that the stream is disturbed at a distance of 10 $R_{wd}$. In this diagram, the portion of the stream absorbing the X-rays produced at the white dwarf surface would need to be the part relatively close to the white dwarf surface. This section of the stream is compressed by the magnetic field lines and has a smaller radius than the broadened portion around the disruption point.

It is instead possible that the broader region of the stream produces the intensity dip. This would require either stream disruption at a larger distance from the white dwarf, at around 20 $R_{wd}$, or a more complex coupling region which further broadens the stream. It is not possible from our observations to distinguish between models predicting quadrupolar fields for the white dwarf and models which allow dipolar fields but delay stream coupling to the field lines until far beyond the white dwarf magnetosphere.

The R-band photometry of BWF93 show a two phased eclipse ingress consisting of a slow decrease in flux lasting approximately 40 sec, attributed to the eclipse of the white dwarf photosphere (BWF93), and a rapid decrease in flux lasting only a few seconds. However, after the end of these decreases, a small but continuous drop is visible in the R-band data until mid-eclipse. This third phase of flux decrease may be due to the eclipse of the accretion stream by the secondary star. The phase of occurrence and duration of this drop are similar to such occultations in other polars including the recently discovered eclipsing polar RE2107-05 (Hakala et al. 1993, Glenn et al. 1994). In Figure 8, the model exhibiting disruption of the accretion stream at a distance of 10 $R_{wd}$ shows that the accretion stream will be completely occulted by the secondary slightly after mid-eclipse. Increasing the coupling region to 20 $R_{wd}$ will produce the end of the stream occultation just at mid-eclipse.

## 8. Conclusions

Several conclusions have been reached in this paper:

(1) The temperature of the soft X-ray region on the surface of the white dwarf has been constrained with the emission arising from a fraction of the white dwarf's surface much smaller than the previously observed cyclotron and UV blackbody emission regions.

(2) The distance to DP Leo has been constrained by the photoelectric absorption to be $\lesssim 500$ pc. An estimate based upon the magnitude of the secondary provides a distance of $260^{+150}_{-100}$ pc.

(3) An intensity dip is observed prior to eclipse. The likely origin is absorption of the emission region on the surface of the white dwarf by an accretion stream which is varying in its position, relative to the secondary, with time.

(4) The X-ray light curve is highly variable showing no emission outside of the bright phase, sinusoidal or little emission prior to the intensity dip, and flare-like emission after eclipse. This suggests that the soft X-ray accretion region may be inhomogeneous with respect to accretion rate per unit area.

(5) X-ray emission was occurring at only one pole in the ROSAT data from 1992. Our modeling suggests, however, that two pole X-ray emission was likely occurring at the time of the EINSTEIN observations in 1979.

(6) A soft X-ray excess exists showing that the soft X-rays are produced primarily from a mechanism other than the reprocessing of bremsstrahlung emission. The great variability observed in soft X-rays, not seen in optical light curves, suggests that the origin is more likely filament accretion.

(7) The magnetic pole of the white dwarf currently rotates slightly faster than the orbital period. This motion may be produced by an oscillation in the position of the magnetic pole of the white dwarf with respect to the secondary star. In addition, the eclipse period observed in DP Leo is longer than the actual orbital period due to the accretion region longitude variations. Magnetic activity induced distortion of the secondary star may be producing both oscillations in the orbital period and the difference between the orbital and white dwarf rotation periods.

(8) A model for the accretion stream is developed assuming initially a ballistic trajectory for the stream and disrupting the stream flow close to the white dwarf surface. This model explains the structure of red light emission within the eclipse.

Models of the magnetic fields in polars may be constrained in the future by determining the shape of the accretion flow close to the white dwarf surface perhaps through combined X-ray, optical emission line and cyclotron emission observations of high inclination systems. Mechanisms producing the complex magnetic fields observed in cataclysmic variables and single white dwarfs may be constrained through the determination of the field structures. In addition, a slightly longer observational baseline is needed to constrain whether the origin of the asynchronous rotation of the white dwarf is produced by orbital period variations or through oscillations in the position of the magnetic pole.

We appreciate helpful discussions with J. Applegate, M. Cropper, J. Nousek and G. Stringfellow. Information on eclipse timings were kindly provided by J. Bailey prior to their publication. C.R.R. acknowledges the support of the Zaccheus Daniels and NASA Life Sciences fellowships. Support was also provided through NASA grant NAG5-1763.

---

Figure 1—Projections of the cyclotron emission regions for DP Leo as determined by CW93 in (a) magnetic coordinates, assuming a centered dipolar field, and (b) rotational coordinates. The separations and field strengths of these emission regions suggest that the field configuration is more complex than that of a centered dipole. Modeling of the X-ray emission observed with EINSTEIN suggests that two X-ray emitting poles existed in 1979 and their locations were near these cyclotron regions. Our ROSAT observations and other optical observations from 1992 suggest that the location of the emission has changed with time. This figure corrects the inconsistent longitude scale in CW93.

Figure 2—ROSAT spectrum (crosses) of DP Leo with blackbody model (solid line). The two lowest energy channels were not used in determining this fit (see text). Severe constraints were placed upon the flux from any harder X-ray component.

Figure 3—ROSAT blackbody model confidence contours. Note the low absorption column density found for this system.

Figure 4—ROSAT light curves and hardness ratios for DP Leo. The top plot combines all of the ROSAT observations while the lower three show individual orbits. The first dip, centered at phase 0.947, is the occultation by the accretion stream while the dip centered at phase 1.00 is the eclipse of the emission region on the white dwarf by the secondary star. Great variability outside of eclipse can be observed within individual DP Leo orbits and between orbits. Each light curve is binned into 500 bins of size 10.78 seconds.

Figure 5—DP Leo light curves from broadband R photometry (BWF93), ROSAT and EINSTEIN. The dashed lines correspond to count rates of zero. The solid lines are model fits to the data assuming a dense filament accretion region which will produce a sinusoidal oscillations in the light curves. The ROSAT data suggest that a structured accretion region may exist resulting in sinusoidal variations dominating prior to eclipse and individual filament accretion dominating after eclipse.

Figure 6—A comparison of the R-band (circles) and ROSAT (histogram) light curves. The intensity dip before eclipse is obvious in the ROSAT data but is not clearly seen in the red data.

Figure 7—Upper panel (a): Deviation from the linear eclipse ephemeris are shown. The dashed line shows a second order fit to the data. Lower panel (b): Accretion spot longitude positions are plotted against time. The dashed line shows a linear fit to the data.

Figure 8—A model of the accretion stream geometry is presented based upon the constraints provided by our ROSAT analysis. The DP Leo system is shown in a center of mass frame at several different phases with the phases listed to the left of each diagram according to the revised eclipse ephemeris in this paper. The accretion stream is disrupted by a presumed quadrupolar field from the white dwarf. The figure includes the primary and secondary stars, their individual orbits (dashed lines), and the accretion stream. In addition, magnetic field lines in the plane containing the line of centers and the white dwarf's rotation axis are shown. The sizes of the white dwarf, the secondary star and their separations are shown to scale based upon the photometric modeling of BWF93.